\documentclass[utf8]{FrontiersinHarvard} 

\usepackage{url,hyperref,microtype,subcaption}
\usepackage[onehalfspacing]{setspace}
\usepackage{float}

\def\keyFont{\fontsize{8}{11}\helveticabold }
\def\firstAuthorLast{Fisher {et~al.}} 
\def\Authors{E. Fisher\,$^{1}$ and R. Smith\,$^{1,2,*}$}
\def\Address{$^{1}$School of Engineering and Built Environment, Sheffield Hallam University, Sheffield, S1 1WB, UK \\
$^{2}$Laboratory for Nuclear Science at Avery Point, University of Connecticut, Groton, CT 06340-6097, USA}
\def\corrAuthor{Robin Smith}

\def\corrEmail{robin.smith@shu.ac.uk}

\begin{document}
\onecolumn
\firstpage{1}

\title[Travelling S-ant-a optimisation]{Ant Colony Optimisation applied to the Travelling Santa Problem} 

\author[\firstAuthorLast ]{\Authors} 
\address{} 
\correspondance{} 

\extraAuth{}

\maketitle

\begin{abstract}

\section{}

\noindent The hypothetical global delivery schedule of Santa Claus must follow strict rolling night-time windows that vary with the Earth’s rotation and a obey an energy budget that depends on payload size and cruising speed. To design this schedule, the Travelling-Santa Ant-Colony Optimisation framework (TSaP–ACO) was developed. This heuristic framework constructs potential routes via a population of artificial ants that iteratively extend partial paths. Ants make their decisions much like they do in nature, following pheromones left by other ants, but with a degree of permitted exploration. This approach: (i) embeds local darkness feasibility directly into the pheromone heuristic, (ii) seeks to minimise aerodynamic work via a shrinking sleigh cross sectional area, (iii) uses a low-cost ``rogue-ant'' reversal to capture direction-sensitive time-zones, and (iv) tunes leg-specific cruise speeds on the fly. On benchmark sets of 15 and 30 capital cities, the TSaP–ACO eliminates all daylight violations and reduces total work by up to 10\% compared to a distance-only ACO. In a 40-capital-city stress test, it cuts energy use by 88\%, and shortens tour length by around 67\%. Population-first routing emerges naturally from work minimisation (50\% served by leg 11 of 40). These results demonstrate that rolling-window, energy-aware ACO has potential applications more realistic global delivery scenarios.



\tiny
 \keyFont{ \section{Keywords:} Travelling Santa Problem, Travelling-Salesman Problem, Ant-Colony Optimisation, energy-aware routing, high-performance computing} 
\end{abstract}


\section{Introduction}

The classical Travelling Salesman Problem (TSP) asks for the shortest route that visits a set of locations once and returns to the start. Although deceptively simple, it is NP-hard and underpins many real-world logistics tasks (\cite{applegate2011traveling}). The related Travelling Santa Problem (TSaP) revisits this puzzle with a twist: Santa must deliver gifts worldwide only during local hours of darkness to maintain his air of mystery and anonymity. Each destination has a strict night-time window that differs by latitude and longitude, making feasibility of a delivery route depend on the Earth’s rotation. These rolling darkness windows convert a purely spatial problem into a spatio-temporal routing challenge; distance-only optimisations inevitably lead to unacceptable daylight deliveries.

Time-window routing is important in transport and logistics pertaining to many areas such as agriculture (\cite{wu2022research}), healthcare (\cite{liu2013heuristic}) and waste collection (\cite{kim2006waste}). As such, this problem has been studied for decades (\cite{solomon1987algorithms}), but the previously published work typically assumes fixed or regionally synchronised windows. However, global delivery schedules are, in some cases, restricted to sending or receiving goods during typical local working hours. Hence, the determination of optimal delivery routes that explicitly incorporate a moving night time window is important. To meet the net-zero goals of governments around the world, ``green'' effective route-finding logistics now optimise distance, speed and load to minimise energy consumption and emissions (\cite{kara2007energy}). Therefore, handling both timing and energy constraints calls for a flexible approach.

Early exact solvers and heuristics for the TSP laid essential groundwork (\cite{applegate2011traveling}). Incorporating service intervals produces the Vehicle Routing Problem with Time Windows (VRPTW), first formalised by \cite{solomon1987algorithms}. Exact branch-and-bound methods solve small VRPTW problems, but meta-heuristics dominate at realistic scales (\cite{braysy2005vehicle}). Most VRPTW research assumes time windows on a single global clock; the rolling night-time constraint of TSaP is largely unexplored. Approaches to NP-hard routing such as the Travelling Santa Problem typically include metaheuristics like evolutionary algorithms (\cite{almufti2023overview}) and tabu search (\cite{barbarosoglu1999tabu}), deep learning methods such as reinforcement learning (\cite{szepesvari2022algorithms}), and hybrid approaches that combine multiple techniques (\cite{li2024bridging}). These methods aim to find near-optimal solutions quickly for problems that are computationally too complex for exact algorithms to tackle.

General purpose meta-heuristics, genetic algorithms, simulated annealing and others, offer robust VRPTW solutions (\cite{gendreau2010solving}). Among swarm techniques, Ant Colony Optimisation (ACO) is notable for its distributed search and positive feedback via pheromone trails (\cite{dorigo1997ant}). The Multiple Ant Colony System extended ACO to VRPTW by separating routing and scheduling colonies (\cite{gambardella1999macs}), yet still assumed synchronous time windows. Only a handful of novelty projects have addressed the TSP directly, such as \cite{strutz2021travelling}, which, on a million-node Christmas tour, minimised distance only, ignoring night-time feasibility. No prior study jointly addresses (i) worldwide darkness windows, (ii) load-dependent energy expenditure, and (iii) a single-vehicle schedule.

This brief research report presents a preliminary Ant Colony Optimisation (ACO) framework that constructs night-feasible, energy-aware Santa Claus deliver routes that visit capital cities across the globe. The main contributions are: (i) integration of rolling darkness windows into pheromone-guided search; (ii) a work based cost function that scales with sleigh load; and (iii) an adaptive cruise-speed model reflecting realistic travel times. The Materials and Methods section surveys related work,  formalises the TSaP, and details the Ant-Colony Optimisation framework. The Results section critiques the performance of preliminary optimised routes, explores convergence behaviour, and quantifies energy savings. The discussion section further highlights real-world applications of this technique and advances that are needed for application to more complex delivery systems.



\section{Materials and methods}
\label{sec:matmethods}


\subsection{Formalising the problem}

Santa’s delivery task can be expressed as vertexes, labelled as $V = \left\{0, 1, . . . , N, 0 \right\}$, where vertex 0 represents Santa’s base at the North Pole. and each $i > 0$ is a delivery location. To reduce the complexity of the problem, each vertex corresponds to a world capital city. Each of these is characterised by geographic coordinates $(\phi_i,\lambda_i)$ expressed in degrees, a population $P_i$, and a sunrise and sunset, in local time, $[t_i^{dusk} , t_i^{dawn}]$, defined on 24–25 December 2025. The night time windows were shortened by a 15-minute buffer. To reduce the computational cost of the problem to a level calculable using a laptop computer, the number of cities chosen was set to a maximum of 40 (one fifth of all world capitals) for this scoping study.

A tour ($\pi$) is an ordered list of edges that: (i) starts and ends at 0, (ii) visits every city exactly once, and (iii) reaches each city during its darkness window. To evaluate a tour, two cost components were introduced:

\noindent \emph{(1) \textbf{Aerodynamic work}} $W_{ij}$ incurred while travelling from city $i$ to city $j$

\noindent \emph{(2) \textbf{A daylight penalty}} $\Omega$ that activates whenever the sleigh
      lands in daylight.
      This is done by introducing an indicator variable
      \[
        \gamma_{ij} \;=\;
        \begin{cases}
        1, & \text{if arrival at }j\text{ is in daylight},\\
        0, & \text{otherwise}.
        \end{cases}
      \]

The constant $\Omega=2.1\times10^{100}$J was chosen to outweigh any plausible work term,  making daylight arrivals hideously unattractive. The total effective tour ``energy'' is therefore defined as

\begin{equation}\label{eq:objective}
  J(\pi)\;=\;
  \sum_{(i,j)\in\pi}\Bigl[\,W_{ij}\;+\;\gamma_{ij}\,\Omega\,\Bigr],
\end{equation}

\noindent and the Travelling Santa Problem seeks the tour ($\pi$) that minimises $J(\pi)$ subject to the feasibility rules above.


\subsection{Classic Ant Colony Optimisation}\label{sec:method:aco-classic}

The power of biological systems in addressing path finding problems is no secret, with well-popularised systems utilising slime mould-inspired algorithms \cite{tero2007mathematical}. With a similar basic philosophy, the Ant Colony Optimisation (ACO) framework optimises a route by simulating tours of a population of artificial ants that iteratively extend partial paths. These digital ants make their decisions much like they do in nature, by following pheromones left previously by other ants, hile maintaining a degree of exploration. This is incorporated into the algorithm during the construction of the tour.
At the construction step at time $t$, an individual ant ($k$) is located at a city ($i$) with a set
$\mathcal{N}_i^{(k)}$ of yet-unvisited cities.  The probability of this ant choosing the next city $j$, is calculated as

\begin{equation}\label{eq:prob}
p_{ij}^{(k)}(t)=
\frac{\tau_{ij}^{\alpha}(t)\,\eta_{ij}^{\beta}}
     {\displaystyle\sum_{l\in\mathcal{N}_i^{(k)}}
       \tau_{il}^{\alpha}(t)\,\eta_{il}^{\beta}},
\end{equation}

Here, $\tau_{ij}(t)$ is the pheromone level on path $(i,j)$ at time $t$ and $\eta_{ij}=1/J_{\pi}$ is the heuristic, using the inverse of the effective work for that part of the journey (so lower effective work increases probability). The $\alpha$ and $\beta$ are powers ($> 0$) that weight the influence of the pheromone against heuristic information. In the zeroth iteration, equal pheromone levels are applied to all paths, a total of 75 ants set out on their journeys and complete their tours.

In the next iteration, the ant pheromone levels along each possible path are updated to weight the choices of the next set of ants, such that more favourable paths are chosen. This is achieved with two updates.

\noindent \emph{(1)\textbf{ Uniform evaporation}} reduces every trail by a factor
$(1-\rho)$, where $\rho\in(0,1)$ is the evaporation rate.

\noindent \emph{(2) \textbf{Deposit}} adds new pheromone in proportion to tour quality.

\noindent Combined, these two effects modify the pheromone level as

\begin{equation}\label{eq:deposit}
\tau_{ij}(t{+}1)=
(1-\rho)\,\tau_{ij}(t)
\;+\;
\sum_{k=1}^{m}\Delta\tau_{ij}^{(k)}(t),
\end{equation}

with

\begin{equation}\label{eq:delta}
\Delta\tau_{ij}^{(k)}(t)=
  \begin{cases}
    Q/J_k, & \text{if edge }(i,j)\text{ belongs to ant }k\text{'s tour},\\[4pt]
    0, & \text{otherwise},
  \end{cases}
\end{equation}\\

Here, $J_k$ is the total expended energy during ant $k$’s tour
(Eq.~\eqref{eq:objective}) and $Q$ is a user-defined deposit factor.
The best tour found so far receives an extra reinforcement pass, biasing
future ants towards high-quality edges. The level at which to weight preferential edges with pheromone level is a delicate balance between following the gained wisdom of earlier ants but also allowing further exploration.

\subsection{Travelling-Santa Extensions}\label{sec:method:aco}
\noindent \textbf{Dynamic cruise speeds}

Each ant begins iteration 0 with a default cruise speed $\bar v_{\text{default}}=7\,650\;$km\,h$^{-1}$, which was chosen as a sufficient enough value to complete a na{\"i}ve, distance-only world tour in 24 hours. At at departure time $t_{\text{dep}}$ (in UTC), before committing to an edge $(i,j)$, the algorithm checks whether the earliest‐possible arrival under this speed would violate the local darkness window at city $j$. The arrival time is computed as

  \[
      t_{\text{arr}}(i,j)
      = t_{\text{dep}} + \frac{d_{ij}}{\bar v_{\text{default}}}.
  \]

  \noindent If $t_{\text{arr}}$ lies within the hours of darkness at destination $j$, the ant may proceed as is. However, if $t_{\text{arr}}$ lies
  before dusk or after dawn, a \emph{feasible cruise speed} that just meets the nearest boundary is computed as
  \[
    v_{ij}(t)=
      \begin{cases}
        \displaystyle
        \frac{d_{ij}}
             {\,t^{\text{dusk}}_{j}(\mathrm{UTC}) - t_{\text{dep}}}, &
            t_{\text{dep}}<t^{\text{dusk}}_{j}\\[8pt]
        \displaystyle
        \frac{d_{ij}}
             {\,t^{\text{dawn}}_{j}(\mathrm{UTC}) - t_{\text{dep}}}, &
            t_{\text{dep}}\ge t^{\text{dusk}}_{j}.
      \end{cases}
  \]
  
 To keep the model physically plausible, speeds are restricted to
  $v_{\min}=10$ km\,h$^{-1}$ and $v_{\max}=15\,000$ km\,h$^{-1}$. An object will burn up in the atmosphere due to extreme heating from air compression and friction at supersonic speeds, with meteors often vaporizing at $24\,000\;$km\,h$^{-1}$ \cite{romig1965physics}. Therefore, the sleigh's maximum velocity was chosen to be sufficiently below this limit. While even slower speeds above $2\,400\;$km\,h$^{-1}$ can ignite flammable materials like clothing, it is not possible for Santa to complete his delivery schedule at such low speeds, so it is assumed that his sleigh and reindeer possess some kind of advanced heat shield.
  If the required speed exceeds $v_{\max}$ or the remaining darkness window
  is negative, the edge is marked infeasible and receives a costly penalty in the heuristic.

  Every 1\% of iterations, $v_{\text{default}}$ is further refined, since from an aerodynamic work perspective, slower speeds are preferable. 
  The program calculates $v_{\text{avg}}$ $-$ the mean cruise speed of the current best tour $-$ and a smoothing update is then applied as $\bar v_{\text{new}} = 0.1\,v_{\text{avg}} + 0.9\,\bar v_{\text{default}}$. This $\bar v_{\text{new}}$ as the reduced default speed is adopted \emph{only} if that tour remains fully night‑feasible. This feedback loop tightens travel‑time estimates without destabilising pheromone learning.\\

\noindent \textbf{Aerodynamic work and payload-driven cross-section}

  Like in many simplified physics calculations, here Santa's sleigh is modelled as a perfect sphere. The sleigh’s energy consumption while travelling between cities $(i,j)$ is
  modelled as aerodynamic work using the formula:
  \[
    W_{ij}=\tfrac12\rho\,A(t)\,v_{ij}^{2}(t)\,d_{ij},
  \]
  where $\rho=1.225\,$kg\,m$^{-3}$ is the density of air at sea level,
  $d_{ij}$ is the geodesic distance in kilometres, and $v_{ij}(t)$ is the
  cruise speed selected by the dynamic-speed module, described earlier.  
  The cross sectional area of Santa's sleigh, $A(t)$, decreases as presents are delivered, reflecting a
  decreasing payload. As presents are delivered, his sack's cross sectional are decreases with the equation
  \[
    A(t)=0.01+0.99\bigl[1-\sum_{k\in\text{visited}}P_k/P_{\text{tot}}\bigr].
  \]
  The $P_k/P_{\text{tot}}$ is the fraction of gifts delivered upon visiting a location. The constant term (0.01m$^2$) prevents the area from collapsing to zero
  after the final present has been delivered.  
  This work term feeds directly into the ACO heuristic through equation \eqref{eq:objective}, biasing ants toward energy-efficient segments. It encourages low-velocity routes that service high-population cities early on in the journey, radpidly lowering the size of the payload, and hence, lowering aerodynamic drag.\\

\noindent \textbf{Adaptive exploration}

  The degree of ant exploration balances the exploitation of strong pheromone
  trails with occasional exploration of random moves. A ``$\varepsilon$‑greedy'' policy selects a random feasible city with probability $\varepsilon_r$ or chooses the city with the highest transition probability (Eq.~\eqref{eq:prob}) otherwise. The pheromone decay schedule maintains exploration early on while encouraging exploitation later \cite{liu2013heuristic}. 
  The exploration probability begins at $\varepsilon_0=0.40$ and decays as the number of iterations increase. A baseline value of $\varepsilon_{\min}=0.05$ makes sure that even during later iterations, there is still a chance for some exploration, rather than just following the pheromone trails. To ensure that the exploration probability first meets $\varepsilon_{\min}$ exactly at the final iteration $R$, the per-step decay factor, $d_{\text{tuned}}$ was set to
      \[
        d_{\text{tuned}}
        \;=\;
        \bigl(\varepsilon_{\min}/\varepsilon_0\bigr)^{1/R}.
      \]

\noindent \textbf{Rogue-ant re-evaluation}\label{sec:rogue}

The darkness constraint makes the cost surface direction-sensitive;
a tour that is perfectly feasible West-to-East may incur heavy daylight
penalties when travelling East-to-West and vice-versa.  
To discover this, each completed route is subjected to a
\emph{rogue-ant check}. Once an ant finishes its forward tour \(\pi^{\text{fwd}}\) a “rogue” copy
      instantly computes the reversed sequence
      \(\pi^{\text{rev}} = (0,\,v_{N},\dots,v_{1},0)\). Both directions are simulated with the same departure time,
      calculating energies \(J(\pi^{\text{fwd}})\) and \(J(\pi^{\text{rev}})\)
      using Eq.~\eqref{eq:objective}. The direction with the lower cost is retained for pheromone deposit,
      while the more expensive one is discarded.

Because reversal costs only \(O(N)\) time per ant, the extra computational time is
negligible relative to the full tour construction. However, inclusion of the rogue ant was found lead to a more rapid reduction of daylight penalties at early iterations in pilot tests. Intuitively, the additional ``look-behind'' step widens the search neighbourhood without requiring new construction heuristics. It is especially valuable when night windows are long in one hemisphere and short in the other $-$ a pattern that naturally skews edge desirability by direction. The overall model parameters used in the presented calculations are given in table \ref{tab:params}.

\begin{table}[htbp]
\caption{The model parameters used in the presented calculations.}
\label{tab:params}
\centering
\begin{tabular*}{\linewidth}{@{\extracolsep{\fill}}lcl}
\toprule
Parameter & Value & Note\\
\midrule
Iterations $R$             & 5\,000 & Number of iterations\\
Ants $m$                   & 75     & Colony size\\
$\alpha,\beta$             & 3,\;2  & Pheromone vs.\ heuristic bias\\
$Q$                        & 1.0    & Deposit factor\\
$\tau_{\min},\tau_{\max}$  & 0.1,\;10 & Pheromone limits\\
$\varepsilon$              & 0.40 $\searrow$ 0.05 & Exploration schedule\\
$v_{\min},v_{\max}$        & 10,\;15 000 km h$^{-1}$ & Cruise-speed bounds\\
Penalty $\Omega$           & $2.1{\times}10^{100}$ J & Daylight penalty\\
\end{tabular*}
\end{table}


\section{Results}

\subsection{Comparative Analysis of Traditional ACO vs TSaP}\label{sec:TSap v aco}

The performance of the full TSaP-ACO approach described in section \ref{sec:matmethods} was measured against the traditional distance-only ACO approach. The convergence behaviour and final energy outcomes of the two algorithms were compared across varying subsets of world capital cities ($N=15, 30, 45$) during a 3000-iteration testing phase.

Convergence was measured by plotting total energy expenditure of a route \eqref{eq:objective} as a function of the number of iterations. Both approaches appear to converge within approximately 500 iterations for $N=15$, with incremental improvements becoming negligible from then on. Notably, the TSaP-ACO achieves lower final energies sooner than the distance-only method, as would be expected, demonstrating its ability to consider additional constraints like aerodynamic drag and rolling darkness windows.


The total energy expenditure was calculated for the traditional distance-only ACO and the full TSaP-ACO approach over several trials. The full TSaP-ACO demonstrated not only a lower median energy expenditure but also a narrower distribution.

Convergence curves reveal how each algorithm copes with growing spatial and temporal complexity. For the smallest instance (\(N=15\)) the trajectories of the distance-only and full TSaP-ACO heuristics are almost identical until roughly iteration~400.  After that point the full model edges ahead by a few percent and both methods stabilise before iteration~1\,000.  

A different story is told for \(N=30\). Here the two curves diverge much earlier, crossing at around iteration~800 and the distance-only run continues to exhibit low-amplitude oscillations well past iteration~2\,500, while the full TSaP-ACO has already stabilised at a lower energy by around 1\,200 iterations. These results suggest that the richer pheromone landscape created by the constraints help
the ants to identify high-quality edges sooner, and that this advantage
grows with the number of cities.

\subsection{Large-Scale Validation: 40–Capital Instance}
\label{sec:exp40}

\begin{figure}[H]
  \centering
  \includegraphics[width=0.75\textwidth]{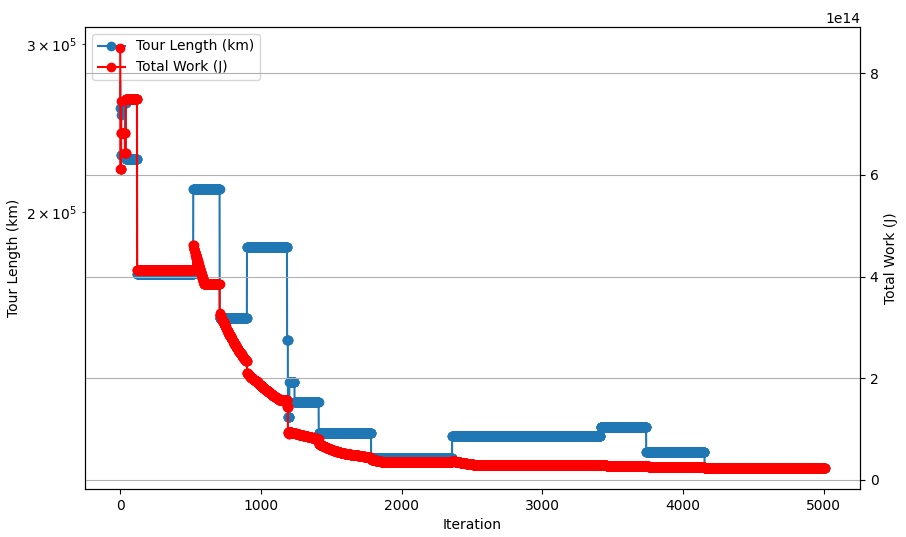}
  \caption{The convergence of distance travelled and ``energy'' expenditure for the 40-capital-city calculation over
           \(6\,000\) iterations.}
  \label{fig:dual_conv_40}
\end{figure}

A final stress-test was run on a set of \(N = 40\) national capitals drawn from all inhabited continents except Antarctica. Figure~\ref{fig:dual_conv_40} overlays tour length and total work on a common timeline. Distinct ``stairs’’ are visible, each marking the discovery of a more efficient valid tour. Over the run, the objective
falls from \(8.5\times10^{14}\,\text{J}\) to
\(2.0\times10^{13}\,\text{J}\), while the tour length shrinks by 67\%. It is worth noting here that a valid near-optimal tour length is found after 1\,100 iterations. However, after that point, the energy expenditure drops significantly, further emphasising that the energy efficiency of a route is not just dictated by length.

\begin{figure}[H]
  \centering
  \includegraphics[width=0.75\textwidth]{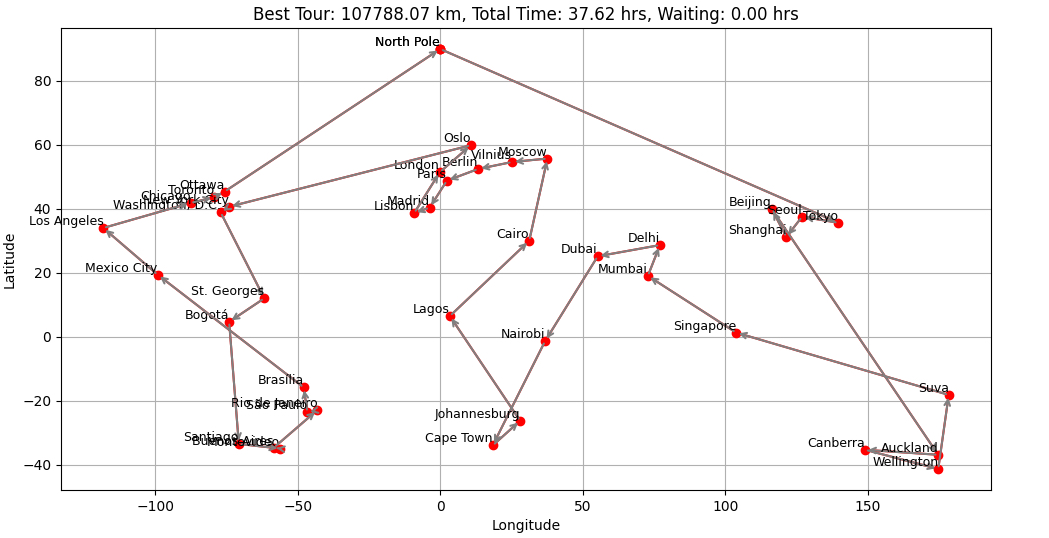}
  \caption{The great-circle view of the optimum 40-city route as determined after \(5\,000\) iterations.}
  \label{fig:best_tour_40}
\end{figure}

The best delivery route is shown in Fig.~\ref{fig:best_tour_40}. It covers \(124\,539\) km in \(31.30\) hours with zero daylight violations. In general, the route tracks East-West, following the Earth's rotation, with some exceptions, which appear to facilitate early visits to high-population cities.

\begin{figure}[H]
  \centering
  \includegraphics[width=0.75\textwidth]{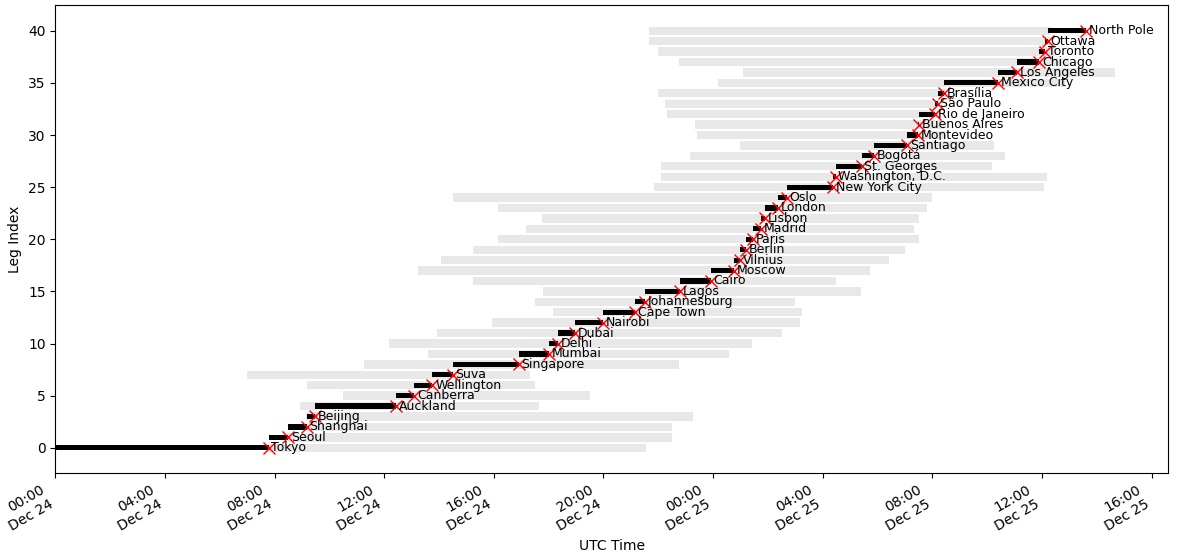}
  \caption{The dark-window utilisation of the 40-city solution.}
  \label{fig:gantt40}
\end{figure}

An alternative way to visualise the journey is depicted in figure~\ref{fig:gantt40}. Here, time runs along the horizontal axis and the index of each stop is on the vertical axis. The night-time windows of each considered capital city are shown by the grey rectangles. Santa's route is shown by the black line and reveals a near–diagonal progression across the grey bars. This demonstrates that the
darkness windows line up in a rough chronological order as the leg index increases. It is also worth noting that at the start of Santa's route, he arrives in the early hours of the evening, whereas after the 
 Shanghai-to-Singapore step he then seems to progress into the latter stages of night, before finishing his tour shortly before sunrise in both North and South America.

\subsection{Load, speed and work distribution}

By tracking the per-leg energy expenditure of this route, it was seen that the first ten delivery stops account for the majority of energy expenditure. This is as expected as the sack of presents is largest at the beginning of the journey and subject to larger aerodynamic drag. Further analysis of the route confirms the expected ``heaviest-first'' delivery policy:
50\% of the entire population are served by leg 11 of 40. This is beneficial as when delivering to the larger-population cities first, Santa's sleigh will have a smaller cross sectional area and therefore will be subject to less aerodynamic drag. This further highlights the goal of minimising energy expenditure rather than distance travelled.

Cruise speeds settle on a tuned ceiling of
\(\approx 3\,500\,{km\,h^{-1}}\) with a small reduction for the first few legs, \(\approx 1\,000\,{km\,h^{-1}}\). These slowest first leg was found such that arrival in Tokyo (the first stop) happened in the hours of darkness.\\


\section{Discussion}

The present calculations represent a paradigm shift towards \emph{timing} rather than geography in the modern ``green'' Travelling-Santa landscape. Placing rolling darkness windows directly inside the pheromone heuristic removed \emph{all} daylight violations \emph{without} adding any holding patterns. In practical terms, the sleigh never wastes fuel while ``circling till nightfall'', a  weakness that distance-only heuristics posses.

Energy-aware routing delivers a second, possibly more useful benefit.  When aerodynamic work, payload decay and leg-specific cruise-speed tuning are jointly optimised, total work falls significantly while the geometric tour length is cut by two thirds. These gains are because the model continually trades a small increase in trip time for a larger reduction in drag (\(\propto v^{2}\)). The algorithm then uses the saved energy to more greatly align with the moving darkness band due to Earth's rotation.

Despite these successes, Fig.~\ref{fig:gantt40} indicates that the current solution does not represent a true optimal tour in terms of energy minimisation. A better solution would fill a greater portion of each city's darkness window, indicating that Santa is travelling slower and thus expending less energy. The presence of under utilised time windows illustrates potential inefficiencies, which likely stem from implementing a constant default speed across all legs of the journey, which only changes when circumstances demand.

To address this, future work could extend the existing adaptive speed model towards dynamically calculating optimal cruise speeds for each individual leg of the journey. Tailoring speeds per individual leg would facilitate a more efficient distribution of travel times, maximising the utilisation of available darkness windows, and thereby reducing overall aerodynamic work and improving energy performance.

Scalability to larger numbers of cities shows promise. Although the search space expanded by 33\%, the 40-node journey only needed 1.67 times the number of iterations compared to the 30-node case to reach a convergence plateau. This supports the idea that richer pheromone landscapes highlight valuable edges more quickly. Unfortunately, the calculation required to explore routes for the full 195-capital city delivery were beyond the computational scope of the laptop computer used for this work. However, access to larger computer clusters would resolve this issue. One improvement to scalability would be to use a GPU to run the individual ants in parallel, which if well-designed, will in turn make for a much faster algorithm, meaning that even more cities can be added into the framework with relative ease.
 
The framework adopts a useful \textit{heaviest-first} strategy: half the population is served by leg 11. Deliveries to high-population hubs naturally occur early, rapidly reducing sack size and diminishing drag, thereby enhancing energy efficiency. Finally, the \emph{rogue-ant} reversal proves cost-effective.  A single, \(O(N)\) reverse-tour check per ant slashes early daylight penalties and sharpens pheromone clusters along the dominant Europe–Americas corridor, adding little computational cost yet expanding the search neighbourhood.

\section{Rights retention statement}

\noindent For the purpose of open access, the author has applied a Creative Commons Attribution (CC BY) license to any Author Accepted Manuscript version of this paper arising from this submission.

\section*{Code availability}

\noindent Source codes, written in python, are available in this data repository: http://github.com****

\section*{Author contributions}
\noindent 
E.F. wrote the code and developed the methodology, and was supervised by R.S. who co-generated ideas. Both authors contributed equally to writing the manuscript.

\section*{Competing interests }
\noindent
The authors declare no competing interests. 

\section*{Materials \& Correspondence}
\noindent
Corresponding author to which requests should be made is R.S.

\section*{Ethics approval}
\noindent
This work was subject to ethics approval procedures of the lead author's institution.

\bibliographystyle{Frontiers-Harvard} 
\bibliography{test}

@incollection{applegate2011traveling,
  title={The traveling salesman problem: a computational study},
  author={Applegate, David L and Bixby, Robert E and Chv{\'a}tal, Va{\v{s}}ek and Cook, William J},
  booktitle={The Traveling Salesman Problem},
  year={2011},
  publisher={Princeton university press}
}

@article{solomon1987algorithms,
  title={Algorithms for the vehicle routing and scheduling problems with time window constraints},
  author={Solomon, Marius M},
  journal={Operations research},
  volume={35},
  number={2},
  pages={254--265},
  year={1987},
  publisher={Informs}
}

@article{wu2022research,
  title={Research on the time-dependent split delivery green vehicle routing problem for fresh agricultural products with multiple time windows},
  author={Wu, Daqing and Wu, Chenxiang},
  journal={Agriculture},
  volume={12},
  number={6},
  pages={793},
  year={2022},
  publisher={MDPI}
}

@article{liu2013heuristic,
  title={Heuristic algorithms for a vehicle routing problem with simultaneous delivery and pickup and time windows in home health care},
  author={Liu, Ran and Xie, Xiaolan and Augusto, Vincent and Rodriguez, Carlos},
  journal={European journal of operational research},
  volume={230},
  number={3},
  pages={475--486},
  year={2013},
  publisher={Elsevier}
}

@article{kim2006waste,
  title={Waste collection vehicle routing problem with time windows},
  author={Kim, Byung-In and Kim, Seongbae and Sahoo, Surya},
  journal={Computers \& Operations Research},
  volume={33},
  number={12},
  pages={3624--3642},
  year={2006},
  publisher={Elsevier}
}

@inproceedings{kara2007energy,
  title={Energy minimizing vehicle routing problem},
  author={Kara, Imdat and Kara, Bahar Y and Yetis, M Kadri},
  booktitle={International conference on combinatorial optimization and applications},
  pages={62--71},
  year={2007},
  organization={Springer}
}

@article{almufti2023overview,
  title={Overview of metaheuristic algorithms},
  author={Almufti, Saman M and Shaban, A Ahmad and Ali, Z Arif and Ali, R Ismael and Fuente, JA Dela},
  journal={Polaris Global Journal of Scholarly Research and Trends},
  volume={2},
  number={2},
  pages={10--32},
  year={2023}
}

@article{barbarosoglu1999tabu,
  title={A tabu search algorithm for the vehicle routing problem},
  author={Barbarosoglu, Gulay and Ozgur, Demet},
  journal={Computers \& Operations Research},
  volume={26},
  number={3},
  pages={255--270},
  year={1999},
  publisher={Elsevier}
}

@book{szepesvari2022algorithms,
  title={Algorithms for reinforcement learning},
  author={Szepesv{\'a}ri, Csaba},
  year={2022},
  publisher={Springer nature}
}

@article{li2024bridging,
  title={Bridging evolutionary algorithms and reinforcement learning: A comprehensive survey on hybrid algorithms},
  author={Li, Pengyi and Hao, Jianye and Tang, Hongyao and Fu, Xian and Zhen, Yan and Tang, Ke},
  journal={IEEE Transactions on evolutionary computation},
  year={2024},
  publisher={IEEE}
}

@article{braysy2005vehicle,
  title={Vehicle routing problem with time windows, Part II: Metaheuristics},
  author={Br{\"a}ysy, Olli and Gendreau, Michel},
  journal={Transportation science},
  volume={39},
  number={1},
  pages={119--139},
  year={2005},
  publisher={INFORMS}
}

@book{gendreau2010solving,
  title={Solving large-scale vehicle routing problems with time windows: The state-of-the-art},
  author={Gendreau, Michel and Tarantilis, Christos D},
  year={2010},
  publisher={Cirrelt Montreal}
}

@article{dorigo1997ant,
  title={Ant colonies for the travelling salesman problem},
  author={Dorigo, Marco and Gambardella, Luca Maria},
  journal={biosystems},
  volume={43},
  number={2},
  pages={73--81},
  year={1997},
  publisher={Elsevier}
}

@misc{gambardella1999macs,
  title={MACS-VRPTW: A multiple ant colony system for vehicle routing problems with time windows},
  author={Gambardella, Luca Maria and Taillard, {\'E}ric and Agazzi, Giovanni},
  year={1999},
  publisher={Istituto Dalle Molle Di Studi Sull Intelligenza Artificiale}
}

@article{strutz2021travelling,
  title={Travelling santa problem: Optimization of a million-households tour within one hour},
  author={Strutz, Tilo},
  journal={Frontiers in Robotics and AI},
  volume={8},
  pages={652417},
  year={2021},
  publisher={Frontiers Media SA}
}

@article{romig1965physics,
  title={Physics of meteor entry},
  author={Romig, Mary F},
  journal={AIAA Journal},
  volume={3},
  number={3},
  pages={385--394},
  year={1965}
}

@article{tero2007mathematical,
  title={A mathematical model for adaptive transport network in path finding by true slime mold},
  author={Tero, Atsushi and Kobayashi, Ryo and Nakagaki, Toshiyuki},
  journal={Journal of theoretical biology},
  volume={244},
  number={4},
  pages={553--564},
  year={2007},
  publisher={Elsevier}
}


\end{document}